\documentclass[12pt]{article}
\usepackage{amsmath} 
\usepackage{multirow} 
\usepackage{amsfonts}
\usepackage{amssymb,graphics,psfrag}
\usepackage{array,epsfig,multirow,stmaryrd,graphicx,mathtools} 
\usepackage{hyperref}
\usepackage{comment}
\usepackage{color}
\input{colordvi.tex}

\def\hybrid{\topmargin -20pt    \oddsidemargin 0pt
        \headheight 0pt \headsep 0pt
        \textwidth 6.25in       % A4 paper
        \textheight 9 in       % A4 paper
        \marginparwidth .875in
        \parskip 5pt plus 1pt 
          \jot = 1.5ex
   }
\hybrid
\numberwithin{equation}{section}
\numberwithin{table}{section}

\newcommand{\beq}{\begin{equation}}
\newcommand{\eeq}{\end{equation}}
\newcommand{\be}{\begin{equation}}
\newcommand{\ee}{\end{equation}}
\newcommand{\bea}{\begin{eqnarray}}
\newcommand{\eea}{\end{eqnarray}}   
\newcommand{\ben}{\begin{eqnarray*}}
\newcommand{\een}{\end{eqnarray*}}                  
\newcommand{\ba}{\begin{aligned}}
\newcommand{\ea}{\end{aligned}}
\newcommand{\bt}{\begin{tabular}}
\newcommand{\et}{\end{tabular}}
\newcommand{\bc}{\begin{center}}
\newcommand{\ec}{\end{center}}
\newcommand{\ox}{\omega}

\newcommand{\cO}{\mathcal{O}}

\newcommand{\cL}{\mathcal{L}}

\newcommand{\cN}{\mathcal{N}}

\newcommand{\IF}{\text{Im}\, \mathcal{F}}

\newcommand{\nn}{\nonumber}

\newcommand{\cref}{{\bf [check ref]}}

\newcommand{\coker}{\mathrm{coker}\;}
\newcommand{\kernel}{\mathrm{ker}\;}

\newcommand{\GUT}{{\rm GUT}}

\def\IZ{{\mathbb Z}}

\def\IP{{\mathbb P}}

\def\IF{{\mathbb F}}

\psfrag{n1}{$\nu_1$}
\psfrag{n2}{$\nu_2$}
\psfrag{n1'}{$\nu_1'$}
\psfrag{n2'}{$\nu_2'$}
\psfrag{n9}{$\nu_9$}
\psfrag{n10'}{$\nu_{10}'$}
\psfrag{t1}{$\tilde{\nu}_1$}
\psfrag{t2}{$\tilde{\nu}_2$}
\psfrag{t9}{$\tilde{\nu}_9$}
\psfrag{t1'}{$\tilde{\nu}_1'$}
\psfrag{t2'}{$\tilde{\nu}_2'$}
\psfrag{t10'}{$\tilde{\nu}_{10}'$}

\def\blfootnote{\xdef\@thefnmark{}\@footnotetext} 
\long\def\symbolfootnote[#1]#2{\begingroup%
\def\thefootnote{\fnsymbol{footnote}}\footnote[#1]{#2}\endgroup}

\begin{document}

\baselineskip=17pt

\begin{titlepage}
\begin{flushright}
\parbox[t]{1.1 in}
{
KIAS-P11081\\
UT-11-42\\
IPMU11-0201
}
\end{flushright}

\begin{center}

\vspace*{ 1.2cm}

{\large \bf Constraints on GUT 7-brane Topology in F-theory}

\vskip 1.2cm

\begin{center}
 {Hirotaka Hayashi$^1$, Teruhiko Kawano$^2$ and Taizan Watari$^3$ }
\end{center}
\vskip .2cm

{\it $^1$ School of Physics, Korea Institute for Advanced Study, Seoul 130-722, Korea\vspace{.2 cm}

$^2$ Department of Physics, University of Tokyo, Tokyo 113-0033, Japan\vspace{.2 cm}

$^3$ Institute for the Physics and Mathematics of the Universe, University of Tokyo,\\ Chiba 277-8583, Japan} 

 \vspace*{1cm}

\end{center}

\vskip 0.2cm

\begin{center}
\abstract{We study the relation between phenomenological requirements and the topology of the surfaces that GUT 7-branes wrap in F-theory compactifications. In addition to the exotic matter free condition in the hypercharge flux scenario of $SU(5)_{\GUT}$ breaking, we analyze a new condition that comes from a discrete symmetry aligning the contributions to low-energy Yukawa matrices from a number of codimension-three singularity points. We see that the exotic matter free condition excludes Hirzebruch surfaces (except $\IF_0$) as the GUT surface, correcting an existing proof in the literature. We further find that the discrete symmetry for the alignment of the Yukawa matrices excludes del Pezzo surfaces and a rational elliptic surface as the GUT surface. Therefore, some GUT 7-brane surfaces are good for some phenomenological requirements, but sometimes not for others, and this aspect should 
be kept in mind in geometry search in F-theory compactifications.}
\end{center}

\hfill %\today
\end{titlepage}

%\tableofcontents

%\newpage

%%%%%%%%%%%%%%%%%%
%
%   Introduction and Summary
%
%%%%%%%%%%%%%%%%%%

\section{Introduction and Summary}

In recent years, F-theory compactifications have been studied for the construction of the realistic models \cite{arXiv:0803.1194, arXiv:1001.0577, arXiv:1009.3497}. A four dimensional low energy effective field theory realized as a non-Abelian gauge theory on 7-branes wrapping on a surface can be a candidate for a supersymmetric Grand Unified Theory (GUT). Since some of important properties of low-energy effective theory in the
visible sector are determined only by geometry of local neighbourhood along 
the GUT surface in Type IIB / F-theory compactifications \cite{Local}, 
correspondence between geometry around the GUT complex surface and 
low-energy physics can be established to some extent without referring 
to global geometry. 
Some class of complex surfaces with certain topology may even be ruled 
out \cite{Beasley:2008kw}, when various phenomenological requirements
are imposed.
%\kesu{Reference \cite{Beasley:2008kw} (see also \cite{arXiv:0904.1218}) 
% \kesu{have} argued that 
% some phenomenological requirements can set constraints on the topology of the GUT surface where 7-branes wrap. }
Reference \cite{Beasley:2008kw} argued, for example, that the condition
of no exotic matter fields in a hypercharge flux scenario of
$SU(5)_{\GUT}$ breaking can exclude Hirzebruch surfaces for the GUT
7-brane. References \cite{Caltech-0904, Hayashi:2009bt, Cordova:2009fg} also 
made it clear that the low-energy Yukawa matrices receive independent
contributions from enhanced singularity points whose numbers are
determined by intersection numbers on the GUT complex surface.
In this article, we further pursue this program, and report some 
constraints on the topology of GUT complex surfaces following from
phenomenological conditions.

% In this paper, we study the restrictions on the topology of the GUT surface 
% by exploiting a new phenomenological constraint as well as the exotic
% matter free condition from the hypercharge flux scenario of
% $SU(5)_{\GUT}$ breaking. The new constraint comes from a discrete
% symmetry solution \cite{Hayashi:2009bt} to the alignment problem of the
% Yukawa matrices generated from a number of intersection points between
% matter curves. We see how the two conditions restrict the topology of
% the GUT surfaces. 

In section \ref{sec:exotic}, we take a moment to have a brief look at 
the exotic matter free condition in the hypercharge flux scenario of 
$SU(5)_{\GUT}$ breaking.
Requiring that chiral multiplets in the off-diagonal components of the 
$5 \times 5$ matrix of $SU(5)_{\GUT}$ be absent below the Kaluza--Klein 
scale, Ref. \cite{Beasley:2008kw} already concluded that all the
Hirzebruch surfaces $\IF_n$ (except $\IF_0$) are ruled out in this scenario. 
% study the restrictions on the GUT 7-brane topology from the exotic
% matter free condition in the hypercharge flux scenario of $SU(5)_{\GUT}$
% breaking. The hypercharge flux yields chiral multiplets in the
% off-diagonal components of the $5 \times 5$ matrix of
% $SU(5)_{\GUT}$. The absence of their zero modes leads to topological
% conditions the GUT surface \cite{Beasley:2008kw}. 
In fact, there is an error in the calculation. We see that the proof
can still be fixed by {\it fully} exploiting the exotic matter free
condition. 
% we show that all the Hirzebruch surfaces except for $\IF_0$ are 
% indeed excluded by the exotic free condition. 
% %, which corrects the proof in \cite{}. 

In section \ref{sec:flavor}, we move on to study consequences of a new
topological condition motivated by phenomenology. Because the number 
of $E_6$ enhancement points is always even in $SU(5)$ models in F-theory \cite{Hayashi:2009bt, Cordova:2009fg}, 
the low-energy up-type Yukawa matrix consists of contributions from 
two or more $E_6$ type points, and the approximately rank-1 structure 
of individual contributions \cite{Hayashi:2009ge, arXiv:0811.2417} is lost
generically. One of the solutions to this problem is to impose a
discrete symmetry $\Gamma$ \cite{Hayashi:2009bt}, so that the contributions
from multiple enhanced singularity points are aligned and the
approximate rank-1 structure is maintained even after all these 
contributions are summed up. 
This discrete symmetry solution sets constraints on the number 
of enhanced singularity points. 
We find that both del Pezzo surfaces and a rational elliptic surface are
excluded by the constraints when $\Gamma=\IZ_2$ (which can be used as 
$R$-parity simultaneously). We also argue that a less general condition for the number of codimension-three singularity points can eliminate del Pezzo surfaces and a rational elliptic surface for all the discrete symmetry $\Gamma$.

% see the restrictions from a discrete symmetry for the alignment of the Yukawa matrices. In general, we have a number of codimension-three singularities which realizes low-energy Yukawa matrices of the rank ${\rm min}(\# \; {\rm of} \; {\rm intersections}, \# \; {\rm of} \; {\rm generations})$ approximately. In particular, the number of the $E_6$ enhancement points is generically even \cite{Hayashi:2009bt, Cordova:2009fg}. Then, a discrete symmetry $G$ can align the contributions to the low energy Yukawa matrices generated from a number of the codimension-three singularities, and recover the approximate rank one structure \cite{Hayashi:2009bt}.  %and generate the realistic flavor structure. 
% Especially, in the case of $G=\IZ_2$, the discrete symmetry could also work for the solution to the dimension-four proton decay problem as argued in \cite{Tatar:2009jk, Hayashi:2009bt}. In this framework, the number of codimension-three singularity points which generate the up-type and down-type Yukawa couplings should be related to the order of the group $G$ and that leads to new conditions. We find that both del Pezzo surfaces and a rational elliptic surface are excluded by the conditions when $G=\IZ_2$. We also argue that a less general condition for the number of codimension-three singularity points can eliminate del Pezzo surfaces and a rational elliptic surface for all the discrete symmetry $G$.

To summarize, 
the $SU(5)_{\GUT}$ breaking by the hypercharge flux excludes Hirzebruch
surfaces (except $\IF_0$), and the discrete symmetry solution for the
alignment problem of the Yukawa matrices excludes del Pezzo surfaces and
a rational elliptic surface. In other words, if one chooses Hirzebruch
surfaces as the $SU(5)_{\GUT}$ GUT surface, one cannot use the
hypercharge flux scenario for $SU(5)_{\GUT}$ breaking, and if one
chooses del Pezzo surfaces or a rational elliptic surface as the
$SU(5)_{\GUT}$ GUT surface, one cannot use a discrete symmetry solution
for the alignment problem of the low-energy Yukawa matrices. Note that
those results hold also when one embeds the GUT surface in a global
Calabi--Yau fourfold. The global embedding of the $SU(5)_{\GUT}$ GUT
surface has been recently discussed in \cite{Caltech-0904, arXiv:0906.4672, arXiv:0908.1784}, and the geometry search has been performed
\cite{arXiv:1101.4908, arXiv:1103.3358}. Our results can constrain the
possible choices for the GUT surface in the geometry search if one
relies on either one of [or on both of] the two scenarios.  
 
%In other words, an arbitrary choice for the $SU(5)_{\GUT}$ GUT surfaces does not allow a particular mechanism for the realistic model buildings. Hence, one must pay attention to the choice of the GUT surfaces in the construction of model buildings.  

%%%%%%%%%%%%%%%%%%
%
%   No-go
%
%%%%%%%%%%%%%%%%%%

\section{No go theorems for SU(5) GUT models}

F-theory compactified on an elliptically fibered Calabi--Yau fourfold realizes a four-dimensional $\cN=1$ supersymmetric theory as a low energy effective field theory. When the elliptic fiber degenerates and generates a singularity of a type $G$ at a codimension one locus $S$ in the base manifold $B_3$, a number of 7-branes wrap on the divisor $S$ and realizes a non-Abelian gauge theory with the gauge group $G$ \cite{hep-th/9602022, hep-th/9602114}. Since we are interested in constructing phenomenological models from F-theory compactifications, we focus on $SU(5)_{\GUT}$ for the gauge group $G$. We will argue that phenomenological requirements indeed restrict the possible candidates for the GUT divisor $S$.

\subsection{Exotic matter free conditions in $SU(5)_{\GUT}$ breaking}
\label{sec:exotic}

%We first choose Hirzebruch surfaces $\IF_n$ for $n \geq 0$ for the GUT divisor $S$. Let $f$ and $\sigma$ be the effective curve classes which generate the middle homology of $\IF_n$. They satisfy the intersection numbers 
%\beq 
%f \cdot f = 0, \qquad f \cdot \sigma = 1, \qquad \sigma \cdot \sigma =-n. \label{int_hir}
%\eeq
%The canonical class of $\IF_n$ is 
%\beq
%K_{\IF_n} = -(n+2)f -2\sigma. 
%\eeq 
 
%In order to restrict Hirzebruch surfaces, we consider the absence of exotic matter fields which may arise from breaking the $SU(5)_{\GUT}$ GUT symmetry into the Standard model gauge group $SU(3) \times SU(2) \times U(1)_Y$. 
The $SU(5)_{\GUT}$ gauge group should be broken to obtain the Standard Model gauge group $SU(3) \times SU(2) \times U(1)_Y$. As for the breaking scenario\footnote{Two other $SU(5)$ breaking scenarios were also referred to in \cite{Beasley:2008kw}. One is to use a Wilson line on a GUT divisor $S$ that is not simply connected, and the other is four-dimensional GUT breaking scenario involving $SU(5)_{\rm GUT}$-${\bf adj}$ chiral multiplets in the effective theory.}, we turn on a hypercharge flux $\cL_Y \equiv (L_Y)^{-5/6} \in H^{2}(S, \IZ)$ \cite{GUT-breaking-via-non-surj-flux-F, arXiv:0806.0634} where the charge against $L_Y$ gives the hypercharge for matter fields. %The scenario can also solve the doublet--triplet splitting problem simultaneously \cite{Beasley:2008kw}. 
The flux $\cL_Y$ should satisfy the BPS condition \cite{arXiv:0802.2969, Beasley:2008dc, Hayashi:2008ba}
\beq
\ox \wedge c_1(\cL_Y) = 0,
\label{BPS}
\eeq
where $\ox$ is a K\"ahler form on the GUT divisor $S$. The hypercharge flux $\cL_Y$ yields chiral multiplets in the off-diagonal blocks of the 5 $\times$ 5 matrix of $SU(5)_{\GUT}$. They are in the representation of ${\bf (3, 2)_{-5/6}}$ and ${\bf (\bar{3}, 2)_{+5/6}}$ under $SU(3) \times SU(2) \times U(1)_Y$. %Since there are no matter fields in such a representation in the spectrum of the Standard Model, those chiral multiplets should not remain in the low energy spectrum. 
Since the presence of those matter fields under the GUT scale breaks the $SU(5)$ gauge coupling unification, we require that there are no zero modes for the matter fields in the representation of ${\bf (3, 2)_{-5/6}}$ and ${\bf (\bar{3}, 2)_{+5/6}}$. 
%The numbers of their zero modes have been computed in \cite{Bershadsky:1997zs, Hayashi:2008ba} (see also \cite{arXiv:0802.2969}) from the analysis via Heterotic -- F-theory duality. 
Then, the exotic matter free conditions are \cite{arXiv:0802.2969, Beasley:2008dc, Hayashi:2008ba} 
\bea
h^{0}(S, \cL_Y^{\pm 1}) &=& 0, \qquad h^{1}(S, \cL_Y^{\pm 1}) = 0, \label{exotic1}\\
\kernel d_2 &=& 0, \qquad  \coker d_2 = 0,\label{exotic2}
%\kernel \tilde{d}_2 &=& 0, \qquad  \coker \tilde{d}_2 = 0. \label{exotic3}
\eea
where $\kernel d_2$ and $\coker d_2$ are understood as their dimension. Here, $d_2$ is a map
\beq
d_2\; : \; H^{0}(S, \cL_Y \otimes \cO(K_S)) \rightarrow H^{2}(S, \cL_Y).
\label{map}
\eeq
The case where the map $d_2$ is non-trivial has been discussed in \cite{hep-th/0208104} in the context of Type IIB string theory.
%which appears in the sepctral sequence calculations as $d_2: E_2^{p,q} \rightarrow E_2^{p+2, q-1}$ for $(p,q)=(0,1)$ where $E_2^{p,q}=H^p(S, R^q \pi_{Z\ast}\pi_Z^\ast\cL_Y)$. The map $\tilde{d}_2$ represents a map $d_2$ with $\cL_Y$ replaced with $\cL_Y^{-1}$. 

When the anti-canonical divisor $-K_{S}$ is an effective divisor, $h^0(S, \cO(D))=0$ for a divisor $D$ always implies $h^0(S, \cO(D+K_S)) = h^2(S, \cO(-D)) = 0$ \cite{Beasley:2008dc}. Hence, the exotic free conditions \eqref{exotic1}--\eqref{exotic2} are equivalent to 
\beq
h^{i}(S, \cL_Y^{\pm 1}) = 0\quad{\rm for}\;i=0,\cdots, 2,
\label{exotic_free}
\eeq
if $-K_S \geq 0$. This includes the cases of Hirzebruch surfaces, del
Pezzo surfaces and rational elliptic surfaces. In such
situations, it follows from \eqref{exotic_free} that
\beq
\chi( S, \cL_Y ) = 0, \qquad \chi( S, \cL^{-1}_Y ) = 0.
\label{eq:chi=0}
\eeq

In fact, Ref.~\cite{Beasley:2008kw} has used the conditions \eqref{eq:chi=0} to set constraints on
Hirzebruch surfaces $\IF_n$. %\kesu{From \eqref{exotic_free}, one also
%has topological conditions $\chi(S, \CL_Y^1)-\chi(S, \cL_Y^{-1}) = 0$
%and $\frac12 (\chi(\IF_n, \cL_Y) + \chi(\IF_n, \cL_Y) )=0$. Then,}
In the following combinations, Ref~\cite{Beasley:2008kw} 
has obtained
\bea
0 = \chi(\IF_n, \cL_Y) - \chi(\IF_n, \cL_Y^{-1}) &=& c_1(\IF_n) \cdot c_1(\cL_Y) = b(n+2)+2a-2bn^2?,\label{hir1}\\
0 = \frac12 (\chi(\IF_n, \cL_Y) + \chi(\IF_n, \cL_Y^{-1}) ) &=&
 1 + \frac{1}{2} c_1(\cL_Y)^2 = 1+\frac12(2ab-b^2 n^2)?\label{hir2} 
%&=& 1+\frac12(b^2 n^2 - b^2(n+2))?\label{hir2}
\eea
Here, $\cL_Y$ is chosen as 
\beq
\cL_Y=\cO(af+b\sigma),
\label{hypercharge1}
\eeq
where $a$ and $b$ are integers, and $f$ and $\sigma$ are generators of the effective curves in $\IF_n$ which satisfy the intersection 
\beq 
f \cdot f = 0, \qquad f \cdot \sigma = 1, \qquad \sigma \cdot \sigma =-n. \label{int_hir}
\eeq
Under the results of the calculation \eqref{hir1} and
\eqref{hir2}, only Hirzebruch surfaces $\IF_n$ with $n = 0, 1$ are
allowed. All the other Hirzebruch surfaces, $n \geq 2$, were excluded for
this reason.
% restricted the ``$n$'' of Hirzebruch surfaces to be $n=0, 1$. Hence, most of the Hirzebruch surfaces $\IF_n$ was excluded for this reason.

However, the correct results for the computation are 
\bea
0 &=&\chi(\IF_n, \cL_Y) - \chi(\IF_n, \cL_Y^{-1}) = c_1(\IF_n) \cdot c_1(\cL_Y) = 2a+2b-bn  ,\label{hir3}\\
0 &=& \frac12 (\chi(\IF_n, \cL_Y) + \chi(\IF_n, \cL_Y^{-1}) ) = 1 + \frac{1}{2} c_1(\cL_Y)^2 = 1+\frac12(2ab-b^2 n) , \label{hir4}
\eea
from the intersection numbers \eqref{int_hir}. Then, the only
constraints that we can derive from \eqref{hir3} and \eqref{hir4} are
$n={\rm even}$, $b=\pm 1$ and $a=b (n-2)/2$. Therefore, the conditions \eqref{hir3} and \eqref{hir4} are not enough to set a bound on the ``$n$'' of Hirzebruch surfaces $\IF_n$.

However, we can indeed exclude all the Hirzebruch surfaces except 
for $\IF_0$ by exploiting all of \eqref{exotic_free}, not 
just (\ref{eq:chi=0}). It will be straightforward to see
that\footnote{
One will also arrive at the same conclusion by using the BPS 
condition (\ref{BPS}) with a K\"ahler form strictly in the interior 
of the K\"ahler cone, instead of using the condition 
$h^0(S, \cL^{\pm 1}_Y ) = 0$. 
If the K\"ahler form is on the boundary of the cone, 
at least one has to be careful in discussing the validity of the
Katz--Vafa style effective field theory description on 7+1 dimensions \cite{hep-th/9606086, arXiv:0802.2969, Beasley:2008dc}. } 
\begin{equation}
 h^0( S , \cL_Y) > 0
\end{equation}
for $b = 1$ and $a = (n-2)/2$ for $S = \IF_n$ with an even $n \geq 2$.
After using all of the exotic matter free condition (\ref{exotic_free}), 
\beq
\IF_0\quad {\rm with} \quad \cL_Y = f - \sigma,\quad {\rm or} \quad -f + \sigma
\eeq
is the only solution in the $S= \IF_n$ series in the hypercharge flux 
scenario of $SU(5)_{\GUT}$ breaking.

\subsection{Discrete symmetry for the alignment of Yukawa matrices}
\label{sec:flavor}

As we have already explained in Introduction, the desirable flavor structures can also set constraints on the GUT divisor in a certain scenario. %An approximate rank one Yukawa coupling is generated from a generic $E_6$ or $D_6$ singularity enhancement point \cite{Hayashi:2009ge}. 
If one could achieve a configuration such that there are only one $E_6$ and one $D_6$ singularity points, % which contribute to Yukawa couplings, the desired flavor structures can be generated by the introduction of the bulk three-form flux \cite{Heckman:2008qa, Cecotti:2009zf}. 
the up-type and the down-type Yukawa couplings have an approximate rank one structure \cite{Hayashi:2009ge, arXiv:0811.2417}. In generic F-theory compactifications, however, one can show that the number of $E_6$ singularity points are always even and the approximate rank one structure is generically broken \cite{Hayashi:2009bt, Cordova:2009fg}. Hence, one has to set some constraints which realize a situation where only one Yukawa point dominantly contributes to a Yukawa coupling in a low energy effective field theory. %Then, one has to factorize matter curves and make completely independent wave functions localize along the separated matter curves in order to realize a situation such that only single $E_6$ Yukawa point contributes to the up-type Yukawa coupling. However, the complete separation of zero modes requires a global factorization and the analysis using Higgs bundle is in fact incomplete \cite{Hayashi:2010zp}. In that case, the model buildings require more involved tasks. 

One way to achieve the desired configuration %which does not require any involved analysis of the global factorization. The solution 
is to impose a discrete symmetry globally \cite{Hayashi:2009bt}\footnote{There are two other solutions to recover the approximate rank one Yukawa matrices. One is to factorize matter curves and make completely independent wavefunctions localize along the separated matter curves \cite{Caltech-0904, Hayashi:2009bt, Cordova:2009fg}. Then, one could realize a situation such that only single $E_6$ codimension-three singularity point contributes to the up-type Yukawa coupling. In this case, %one has to make a global factorization in order to realize a complete separation of the zero modes. The global factorization issue has also been discussed in the context of an unbroken $U(1)$ symmetry in low energy effective theories \cite{Hayashi:2010zp, arXiv:1006.0226, arXiv:1006.0483}. 
the matter wavefunctions need to be separated at the level of factorization of spectral surfaces, not at the level of matter curves \cite{hep-th/0602238, Tatar:2009jk}. In models without an $SO(10)$ or $SU(6)$ GUT divisors, this factorization condition needs to be lifted to a constraint on the global 4-fold geometry \cite{Tatar:2009jk, Hayashi:2010zp, arXiv:1006.0226, arXiv:1006.0483}. The other is to tune the complex structure moduli of the matter curve for {\bf 10} representation and make the wavefunctions of {\bf 10} matter fields localize along the matter curve \cite{Hayashi:2009bt}. Then, one codimension-three singularity point makes a dominant contribution to the Yukawa matrix in the four-dimensional effective theory, and recovers the approximate rank one structure of the Yukawa matrices in the Standard Model.}.  %\footnote{In fact, \cite{Hayashi:2009bt} has proposed a different solution which does not need a single Yukawa point. On the contrary to that, the solution makes use of the number of Yukawa points. When the wave functions on each matter curve are localized along the matter curve by tuning certain complex structure moduli of the matter curve, the leading contribution to a Yukawa coupling effectively comes from one Yukawa point and the approximate rank one structure can be realized. In this scenario, the sub-structures of Yukawa couplings are generated from overlaps of the wave functions on the other Yukawa points. When the matter curves are genus one curves, it is necessary to tune only one complex structure modulus. Hence, the desired flavor structures could be realized in a large region of the landscape. The tuning could be ultimately explained from moduli stabilizations. Although the moduli stabilizations may lead to some geometric constraints, it is beyond the scope of this paper and we do not address the issue here.} 
If the discrete symmetry relates all the codimension-three singularity
points of the same type, all the Yukawa matrices generated from the codimension-three singularities are aligned and have the same structures. Hence, one can recover the approximate rank one structure in that situation.  

Introducing a discrete symmetry has another phenomenological
motivation. Imposing $\IZ_2$ symmetry has been considered in
\cite{Tatar:2009jk, Hayashi:2009bt} for a solution for the prohibition
of the renormalizable R-parity violating operators. In order to forbid
the dimension-5 proton decay operators also, the $\IZ_2$ symmetry is not
enough and one has to extend it to a larger discrete symmetry (e.g., \cite{320886}). Therefore, imposing a discrete symmetry may lead to realize the desirable flavor structure and also forbid the dangerous proton decay operators simultaneously. 

Then, let us see how the discrete symmetry for the desirable flavor
structure sets topological constraints on the GUT divisor $S$ of $SU(5)$
GUT models. Let the symmetry group be $\Gamma$. Imagine a case where all the elements of the symmetry group $\Gamma$ acts on the global geometry non-trivially so that they act on
the $E_6$ singularity points [and $D_6$ type points] faithfully and 
transitively. It then follows that
%\kesu{Since the discrete symmetry $G$ should relate all the $E_6$ singularity points or all the $D_6$ singularity points to generate the approximate rank one Yukawa couplings, the requirement for it is }
\beq
\# E_6 = |\Gamma|, \qquad \# D_6 = |\Gamma|.
\label{flavor1}
\eeq 
More generally, there could be a normal subgroup $H_{E_6}$ or $H_{D_6}$
of $\Gamma$ which acts non-trivially on the {\it neighborhood} of the
$E_6$ or $D_6$ singularity points but acts trivially on the {\it points} respectively. In that case, the requirement \eqref{flavor1} is relaxed to
\beq
\# E_6 = \frac{|\Gamma|}{|H_{E_6}|}, \qquad \# D_6 = \frac{|\Gamma|}{|H_{D_6}|}.
\label{flavor2}
\eeq

\subsubsection{Del Pezzo surfaces}

First, we apply the constraint \eqref{flavor1} to del Pezzo surfaces, $dP_{n}$ and $\IP^1 \times \IP^1 (= \IF_0)$, assuming that $H_{E_6} = H_{D_6} = \phi$. %\kesu{In general,} 
The numbers of the $E_6$ and $D_6$ singularity enhancement points on the GUT divisor $S$ are \cite{hep-th/9908193, Hayashi:2008ba}
\beq
\# E_6 = (5K_S + \eta) \cdot (4K_S + \eta), \qquad \# D_6 = (5K_S + \eta) \cdot (3K_S + \eta),
\eeq
where $\eta$ is a divisor in $S$ which satisfies $c_1(N_{S|B_3}) = 6K_S + \eta$. Then, we have 
\beq
0 = \# D_6 - \# E_6 = (5K_{S} + \eta) \cdot (-K_{S}), \label{flavor_dp1}
\eeq
because of \eqref{flavor1}. Since $|5K_{S} + \eta|$ is a class of the matter curve for the {\bf 10} representation, $5K_S + \eta$ is an effective divisor. In the case of del Pezzo surfaces, then, one can see that Eq.~\eqref{flavor_dp1} contradicts the ampleness of $-K_{dP}$. This follows from Nakai's ampleness criterion, which states that the necessary and sufficient conditions for an ample divisor $D$ on a non-singular projective surface $S$ are
\beq
D \cdot D > 0, \qquad C \cdot D > 0 
\eeq
for any curve $C$ in $S$. Therefore, any del Pezzo surfaces do not
satisfy the constraints (\ref{flavor1}), (\ref{flavor_dp1}).

The constraint \eqref{flavor1} is in fact stronger than the general constraint \eqref{flavor2}. However, in the case of $\Gamma=\IZ_2$, one can show that any del Pezzo surfaces cannot satisfy the general constraint \eqref{flavor2}. Note that the number of the $D_6$ singularity points is always larger than the number of the $E_6$ singularity points, since 
\beq
\# D_6 - \# E_6 = (5K_{dP} + \eta) \cdot (-K_{dP}) > 0.
\eeq
Here, we use the fact that $5K_{dP} + \eta$ is an effective divisor and $-K_{dP}$ is an ample divisor. Therefore, the constraint \eqref{flavor2} is satisfied if and only if $\# D_6 = 2$ and $\# E_6 = 1$. However, the number of the $E_6$ singularity points is always even \cite{Hayashi:2009bt, Cordova:2009fg}. Then, the only possibility of $\# D_6 = 2$ and $\# E_6 =1$ cannot be realized. Namely, it is impossible to satisfy the constraint \eqref{flavor2} when $\Gamma = \IZ_2$. %Note that we have only used the fact that $-K_{dP}$ is ample in the argument. Since $-K_{\IF_0}$ is also ample, $\IF_0$ also does not satisfy the constraint \eqref{flavor1}, nor also \eqref{flavor2} in the case of $\Gamma=\IZ_2$.

%\subsection{Rational elliptic surfaces}

%The last example is rational elliptic surfaces. Rational elliptic surfaces are elliptic fibration over $\IP^1$. They can be also thought to be nine points blow up of $\IP^2$. The nine points are the intersection points of two degree three curves in $\IP^2$. Let $H$ be a hyperplane class and $E_i, i=1, \cdots, 9$ are exceptional divisors. The intersection property is the same as \eqref{int_dp}. The generators for the cone of the effective curves in rational elliptic surfaces are 
%\beq
%f = 3H - \sum_{i=1}^{9} E_i, \qquad C_i,
%\eeq
%where $C_i^2 = -1$ and $C_i \cdot f = 1$. $f$ is the elliptic fiber class. In fact, the line bundle \eqref{hypercharge2} also satisfies the BPS condition \eqref{BPS} and the exotic free conditions \eqref{exotic_free} with a specific choice of the K\"ahler class also in the case of rational elliptic surfaces. 

\subsubsection{Rational elliptic surface}

We next apply the constraint \eqref{flavor1} to a rational elliptic surface. Since the anti-canonical divisor of a rational elliptic surface is not ample, one cannot use the discussion of del Pezzo surfaces for the constraints from the discrete symmetry solution in the case of a rational elliptic surface. %First we consider the constraint \eqref{flavor1}. 
The condition \eqref{flavor_dp1} in the case of a rational elliptic surface implies that the matter curve of the class $5K_S + \eta$ is parallel to the elliptic fiber {\bf (i)}. In the case of special complex structure moduli of a rational elliptic surface, the elliptic fiber can degenerate into smaller irreducible pieces at some special points on the base $\IP^1$. In such a case, the matter curve can be one of the components ({\bf ii}). In these cases, the number of $E_6$ singularity points are 
\bea
\# E_6 &=& (5K_S + \eta) \cdot (- 2K_S) + (5K_S + \eta) \cdot (6K_S + \eta), \nonumber \\
&=& {\rm deg}K_{\bar{c}_{{\bf 10}}} = 2g-2, \nn \\
&=& 0 \quad {\rm for}\; {\bf (i)},\\
&{\rm or}& -2 \quad  {\rm for}\; {\bf (ii)},
\eea
where $\bar{c}_{{\bf 10}}$ is a curve in the class 
$|5K_S + \eta|$. 
Hence, neither case realizes the configuration 
$\# E_6 > 0$. 
%\kesu{Namely, a rational elliptic surface does not satisfy the 
%constraint \eqref{flavor1}.} 
The negative result for case {\bf (ii)} means that the condition (\ref{flavor1}) will never be satisfied.  In case {\bf (i)}, the condition (\ref{flavor1}) means that there is no $E_6$ type singularity point in the GUT complex surface $S$, and the up-type Yukawa couplings need to be generated through D-brane instantons \cite{arXiv:0707.1871, arXiv:0811.2936}.

The general constraint \eqref{flavor2} cannot be %}\kesu{is not
%also} 
satisfied in the case of $\Gamma = \IZ_2$ for a rational elliptic 
surface, either. Given the fact that $\# E_6$ is even, 
while $\#E_6 = \#D_6$ cannot be realized, as we have seen above, 
%\kesu{Considering that $\# E_6$ is even and 
%$\# E_6 = \# D_6$ is excluded,} 
the only possibility 
under $\Gamma = \IZ_2$ is $\# E_6 = 2$ and $\# D_6 = 1$. 
This implies that
\beq 
\# D_6 - \# E_6 = (5K_S + \eta) \cdot (-K_{S}) = -1.
\label{flavor3}
\eeq  
Since $5K_{S}+ \eta$ is an effective divisor, the curve class can be written as $a f + \sum_i b_i C_i$ where $a$ and $b_i$ are positive integer. $f$ is the elliptic fiber class and $C_i$'s are curves which satisfy $C_i^2=-1, C_i \cdot f = 1$. Those are the generators of the cone for the effective curves. Due to the intersection number, $(af + \sum_i b_i C_i) \cdot (-K_{S}) = \sum_i b_i > 0$, it is impossible to satisfy \eqref{flavor3}.

\vspace*{1cm}
\noindent
{\bf Acknowledgments}: We would like to thank Kenji Hashimoto for discussions. T.K. and T.W. thanks the organizers of a program ``Branes,
Strings and Black holes'' in YITP, Kyoto University, where a part of the
work was done. This work was supported in part by a Grant-in-Aid
\# 23540286 (T.K.), by a Grant-in-Aid for Scientific Research on Innovative 
Areas 2303 and by WPI Initiative from MEXT, Japan (T.W.).

\end{document}